\documentclass{iaus}
\usepackage{graphicx}

\title[Accurate fundamental parameters of eclipsing binary stars] 
{Accurate fundamental parameters of eclipsing binary stars}

\author[Southworth, Smalley, Maxted and Etzel]
       {J.\ Southworth$^1$, B.\ Smalley$^1$, P.\ F.\ L.\ Maxted$^1$ and P.\ B.\ Etzel$^2$}

\affiliation{$^1$\,Department of Physics and Chemistry, Keele University, Staffordshire, ST5 5BG, UK \break email:
        jkt@astro.keele.ac.uk (JS), pflm@astro.keele.ac.uk (PFLM), bs@astro.keele.ac.uk (BS) \\[\affilskip]
        $^2$\,Department of Astronomy, San Diego State University, San Diego, CA 92182, USA \break email: etzel@mintaka.sdsu.edu}

\pubyear{2004}
\volume{224}
\pagerange{1--15}
\date{??? and in revised form ???}
\jname{The A-Star Puzzle}
\editors{J. Zverko, W. W. Weiss, J. Ziznovsky \& S. J. Adelman, eds}
\begin{document}

\newcommand{\Msun}{\ensuremath{\,{\rm M}_\odot}}                        
\newcommand{\Rsun}{\ensuremath{\,{\rm R}_\odot}}                        
\newcommand{\ion}[2]{{#1}\,{\sc {\small{#2}}}}                          
\newcommand{\Teff}{\ensuremath{T_{\rm eff}}}                            
\newcommand{\logg}{\ensuremath{\log g}}                                 
\newcommand{\FeH}{\ensuremath{\left[\frac{\rm Fe}{\rm H}\right]}}       
\newcommand{\kms}{\,km\,s$^{-1}$}                                       
\newcommand{\OmC}{\ensuremath{O\!\!-\!\!C}}                             
\newcommand{\degr}{\ensuremath{^{\circ}}}                               
\newcommand{\cms}{\,cm\,s$^{-1}$}                                       
\newcommand{\Mbolsun}{\ensuremath{M_{\rm bol \odot}}}                   
\newcommand{\Lsun}{\ensuremath{L_\odot}}                                
\newcommand{\sci}[2]{{#1}\!\times10^{#2}}                               

\maketitle

\begin{abstract}
The study of detached eclipsing binaries is one of the most powerful ways to investigate the properties of individual stars and stellar systems. We present preliminary masses, radii and effective temperatures for the eclipsing binary WW\,Aurigae, which is composed of two metallic-lined A-type stars. We also reanalyse the data on HD\,23642, an A-type eclipsing binary member of the Pleiades open cluster with a metallic-lined component, and determine its distance to be $139 \pm 4$\,pc. This is in agreement with the traditional Pleiades distance, but in disagreement with distance to the Pleiades, and to HD\,23642 itself, derived from Hipparcos trigonometrical parallaxes.
\keywords{stars: binaries: close, stars: binaries: eclipsing, stars: binaries: spectroscopic, stars: chemically peculiar, stars: distances, stars: evolution, stars: fundamental parameters}
\end{abstract}

\firstsection

\section{Introduction}

The analysis of photometric and spectroscopic observations of double-lined detached eclipsing binaries remains one of the most powerful ways to investigate the properties of individual stars. The analysis of light curves and radial velocity curves allows the masses, radii and  surface gravities of two stars to be found to an accuracy of better than 1\%. Further analysis using spectral synthesis or photometric index calibration techniques also allows the derivation of accurate effective temperatures, absolute magnitudes and distances of the components of the eclipsing system (Andersen, 1991; Southworth, Maxted \& Smalley, 2004c). 

Accurate dimensions of a detached eclipsing binary (dEB) provide a discriminating test of theoretical stellar evolutionary models because the two components of the system have the same age and chemical composition but, in general, different masses and radii (Southworth, Maxted \& Smalley, 2004b). The observation and analysis of dEBs in stellar clusters and associations may allow the calculation of accurate dimensions of more than two stars of the same age and initial chemical composition (Southworth, Maxted \& Smalley, 2004a). The distance of the cluster can be found from analysis of an eclipsing member, without the use of theoretical calculations and free of the difficulties of the main-sequence fitting technique. Furthermore, the age, metallicity and helium abundance of the cluster as a whole can be found from comparison of the observed masses and radii of the dEB with theoretical stellar models (Southworth {\it et al}., 2004a).

Another use of dEBs is to investigate the physical processes at work in single stars. This is particularly useful if one or both components belongs to a class of peculiar or poorly understood stars, for example slowly pulsating B stars (e.g., V539\,Arae; Clausen, 1996), or metallic-line stars (e.g., WW\,Camelopardalis; Lacy {\it et al}., 2002). Metallic A\,stars are well represented in the compilation of accurate eclipsing binary data by Andersen (1991) but the physical processes and particular conditions of occurrence are still not fully understood.

\subsection{WW Aurigae}

WW\,Aurigae (HD\,46052, HIP\,31173, $V = 5.9$, period  = 2.52\,d) is a dEB with an accurate Hipparcos parallax distance of $84 \pm 8$\,pc (Perryman, {\it et al}., 1997). Its eclipsing nature was discovered independently by Solviev (1918) and Schwab (1918). Huffer \& Kopal (1951) and Piotrowski \& Serkowski (1956) undertook photoelectric observations but were both hampered by bright skies. Etzel (1975) observed excellent photoelectric light curves in the Str\"omgren $uvby$ filters, consisting of about one thousand observations in each filter. 

Kiyokawa \& Kitamura (1975) published excellent $UBV$ light curves of WW\,Aur and analysed them using a procedure based on rectification (Russell \& Merrill, 1952; Kitamura, 1967). Kitamura, Kim \& Kiyokawa (1976) published good photographic spectra and derived accurate absolute dimensions of WW\,Aur by combining their results with those of Kiyokawa \& Kitamura.

The rotational velocities of the components were found to be 35 and 55\kms\ from CCD spectra by Abt \& Morrell (1995), who also classified the stars as Am (A2,A5,A7) where the bracketed spectral types have been obtained using the \ion{Ca}{II} K line, Balmer lines, and metallic lines.

\subsection{HD\,23642}

HD\,23642 (HIP\,17704, $V = 6.8$, period = 2.46\,d) was discovered to be a double-lined spectroscopic binary by Pearce (1957) and Abt (1958). Its Hipparcos parallax gives a distance of $111 \pm 12$\,pc (Perryman, {\it et al}., 1998). Abt \& Levato (1978) provided a spectral classification of A0\,Vp\,(Si) + Am, where the metallic-line character of the secondary star relies on the presence of strong \ion{Fe}{i} lines. Torres (2003) discovered shallow secondary eclipses in the Hipparcos photometric data of HD\,23642 and also presented an accurate spectroscopic orbit. 

Munari {\it et al}.\ (2004, hereafter M04) derived precise absolute masses and radii of both components from five high-resolution \'echelle spectra, and complete $BV$ photoelectric light curves. From consideration of the effective temperatures and bolometric magnitudes of the components, and using the Wilson-Devinney light curve fitting code (Wilson \& Devinney, 1971; Wilson, 1993) they found the distance to HD\,23642 to be $131.9 \pm 2.1$\,pc, where the quoted error is the formal error of the fit.

HD\,23642 is is an important system because it is a member of the Pleiades open cluster, for which the distance is controversial.

\subsection{The distance to the Pleiades}

The Pleiades is a nearby young open star cluster which has been exhaustively studied by many researchers. The `long' distance scale of $132 \pm 3$\,pc is supported by main sequence fitting analyses and photometric calibrations (e.g., Johnson, 1957; Crawford \& Perry, 1976; Meynet, Mermilliod \& Maeder, 1993) and analysis of the astrometric binary Atlas (Pan, Shao \& Kulkarni, 2004). Trigonometric parallax observations from the ground ($131 \pm 7$\,pc; Gatewood {\it et al}., 2000) and from the Hubble Space Telescope ($135 \pm 3$\,pc; Benedict, 2004) also agree with the `long' distance scale. 

The `short' distance scale of $120 \pm 3$\,pc is derived from Hipparcos parallaxes (van Leeuwen, 2004) but some researchers have suggested that the Hipparcos parallaxes are correlated on small angular scales (Makarov, 2002). However, van\,Leeuwen (1999) has found other nearby open clusters with main sequences as faint as the Pleiades (located with the use of Hipparcos parallaxes). 

Castellani {\it et al}.\ (2002) have shown that current theoretical stellar evolutionary models can fit the Pleiades main sequence if a low metal abundance of $Z = 0.012$ is adopted. However, Stello \& Nissen (2001) used a metallicity-insensitive photometric technique to demonstrate that, if the Hipparcos parallaxes were correct, the main sequence Pleiades stars were implausibly fainter than their counterparts in the field. Also, Boesgaard \& Friel (1990) have measured the iron abundance of the Pleiades to be approximately solar ($\FeH = -0.034 \pm 0.024$) from high-resolution spectra of twelve F dwarfs in the cluster.

Pinsonneault {\it et al}.\ (1998) and Soderblom {\it et al}.\ (1998) investigated this discrepancy in distance determinations by performing a main sequence fitting analysis of the Pleiades and by attempting to discover nearby subluminous stars with solar metal abundance. They concluded that the Hipparcos measurements were most likely wrong and also noted that the most discrepant measurements belong to stars in the centre of the Pleiades, where the stellar density is the highest. This is consistent with the suggestion that the Hipparcos parallaxes are biased where the apparent stellar density is high. Further details can be found in Southworth, Maxted \& Smalley (2004c).


\section{WW Aurigae}

\subsection{Spectroscopic analysis}

A total of 59 grating spectra were obtained in 2002 October using the 2.5\,m Isaac Newton Telescope on La Palma, covering the wavelengths 4220--4500\,\AA\ at a two-pixel resolution of 0.2\,\AA.

Radial velocities were determined from the observed spectra using the two-dimensional cross-correlation algorithm {\sc todcor} (Zucker \& Mazeh, 1994). In this algorithm two template spectra are simultaneously cross-correlated against each observed spectrum for a range of possible velocities for each template. Template spectra of seven standard stars, with spectral types between A0\,IV and F5\,V, were used, with circular spectroscopic orbits being fitting to the radial velocities calculated using each combination of spectra. The final spectroscopic orbital parameters are calculated from the mean and 1\,$\sigma$ uncertainties of the individual good spectroscopic orbits.

\begin{figure} \includegraphics[width=\textwidth,angle=0]{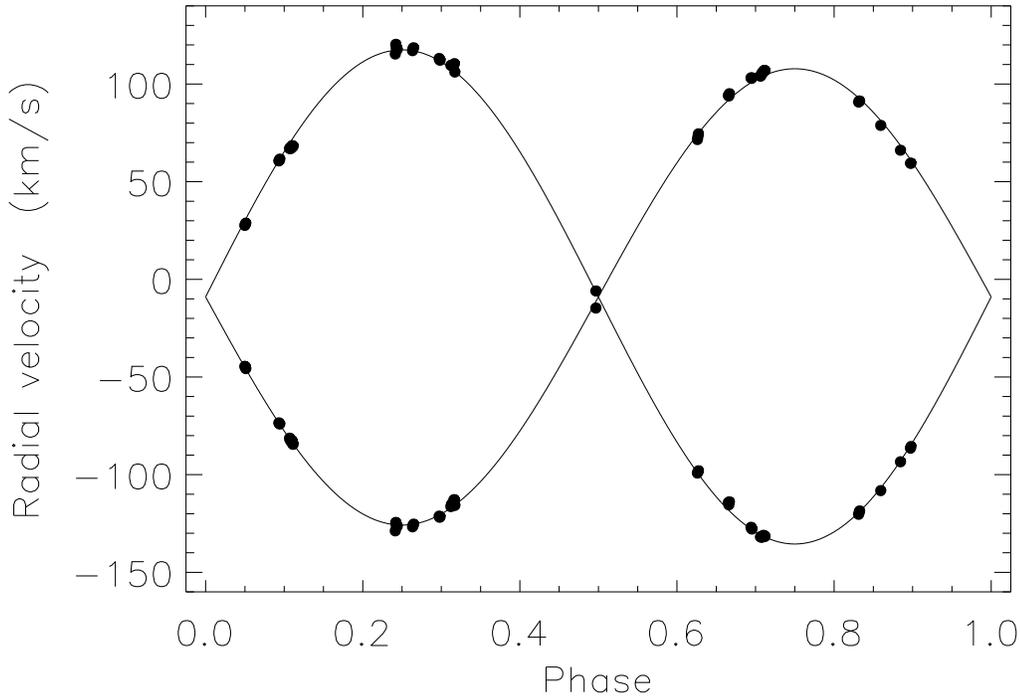} \\ \caption{\label{fig:wwaur:rvorbit} 
Spectroscopic orbit for WW\,Aur. The observed radial velocities are shown with filled circles and the best-fitting spectroscopic orbit is shown using unbroken lines.} \end{figure}

The final spectroscopic orbital parameters are $K_{\rm A} = 116.81 \pm 0.23$\kms, $K_{\rm B} = 126.49 \pm 0.28$\kms, for a circular orbit with an orbital period of 2.52\,d. This gives a mass ratio of $q = 0.9235 \pm 0.0027$ and individual minimum masses of $M_{\rm A} \sin^3 i = 1.9590 \pm 0.0074$\Msun\ and $M_{\rm B} \sin^3 i = 1.8090 \pm 0.0065$\Msun. The best individual spectroscopic orbit has been selected and plotted in Figure~\ref{fig:wwaur:rvorbit}.

\subsection{Light curve analysis}

\begin{figure*} \includegraphics[width=\textwidth,angle=0]{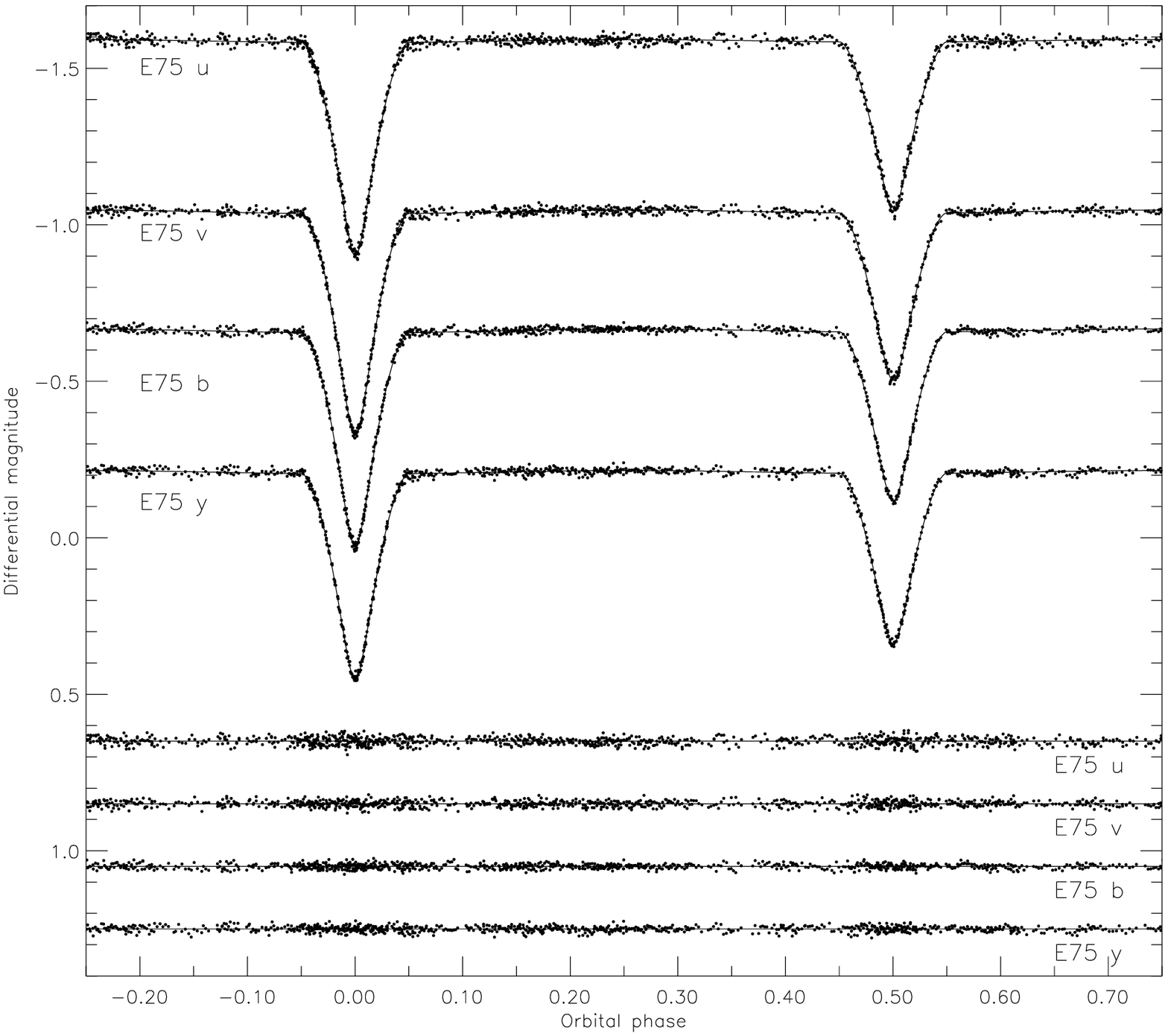} \\ 
\caption{\label{fig:wwaur:lcfit} The Etzel (1975) differential light curves of WW\,Aur, compared to the best-fitting light curves found using {\sc ebop}. The residuals of the fits are plotted with magnitude offsets for clarity} \end{figure*}

The light curves of Etzel (1975), Kiyokawa \& Kitamura (1967) and Huffer \& Kopal (1951) of WW\,Aur were analysed using {\sc ebop} (Nelson \& Davis, 1972; Popper \& Etzel, 1981), a simple and efficient light curve fitting code in which the discs of stars are modelled using biaxial ellipsoids. It is applicable only to eclipsing binaries which are sufficiently well detached for the stellar shapes to be close to spherical, as in the case of WW\,Aur. Investigations showed that the light curves display negligible eccentricity and third light so these quantities were fixed at zero for the final solutions. The light curves and their best fits are plotted in Figure~\ref{fig:wwaur:lcfit}.

Uncertainties in the values of the photometric parameters have been estimated using Monte Carlo simulations (Southworth, Maxted \& Smalley, 2004b). After finding the best fit, the model light curve was evaluated at the phases of the observed datapoints. Random Gaussian noise, of the same size as the standard error of the fit, was added to simulate observational noise. {\sc ebop} was used to fit the resulting `observation-like' light curve, and this process was undertaken ten thousand times for each real light curve.

The $uvby$ photometric indices for the individual components of WW\,Aur are $(b-y) = 0.074 \pm 0.020$ and $0.091 \pm 0.027$, $m_1 = 0.216 \pm 0.036$ and $0.254 \pm 0.047$ and $c_1 = 0.977 \pm 0.039$ and $0.895 \pm 0.0524$ for the primary and secondary stars respectively. The calibration of Moon \& Dworetsky (1985) then gives effective temperatures of $8350 \pm 200$\,K and $8170 \pm 300$\,K for the two stars and indicates that the light of the system suffers from negligible interstellar absorption. If the system value of the H$\beta$ index is assumed to be a reasonable estimate for the individual values for each star, the same results are found and it is confirmed that interstellar absorption is negligible.

\subsection{Absolute dimensions and comparison with stellar models}

\begin{table} \begin{center} 
\begin{tabular}{l r@{\,$\pm$\,}l c r@{\,$\pm$\,}l} \hline
\hspace{200pt}                            & \multicolumn{2}{c}{WW\,Aur A} && \multicolumn{2}{c}{WW\,Aur B} \\ \hline
Mass (\Msun)                              & 1.964     & 0.007     &\hspace{40pt}& 1.814     & 0.007     \\
Radius (\Rsun)                            & 1.980     & 0.009     & & 1.807     & 0.009     \\
Surface gravity \logg\ (\cms)             & 4.160     & 0.007     & & 4.165     & 0.007     \\ 
Effective temperature (K)                 & 8350      & 200       & & 8170      & 300       \\
Equatorial rotational velocity (\kms)     & 35        & 10        & & 55        & 10        \\
Synchronous rotational velocity (\kms)    & 39.69     & 0.18      & & 36.24     & 0.18      \\
\hline \end{tabular} \end{center} \caption{\label{table:wwaur:dimensions} Absolute dimensions and astrophysical parameters for the components of WW\,Aur.} \end{table}

\begin{figure} \includegraphics{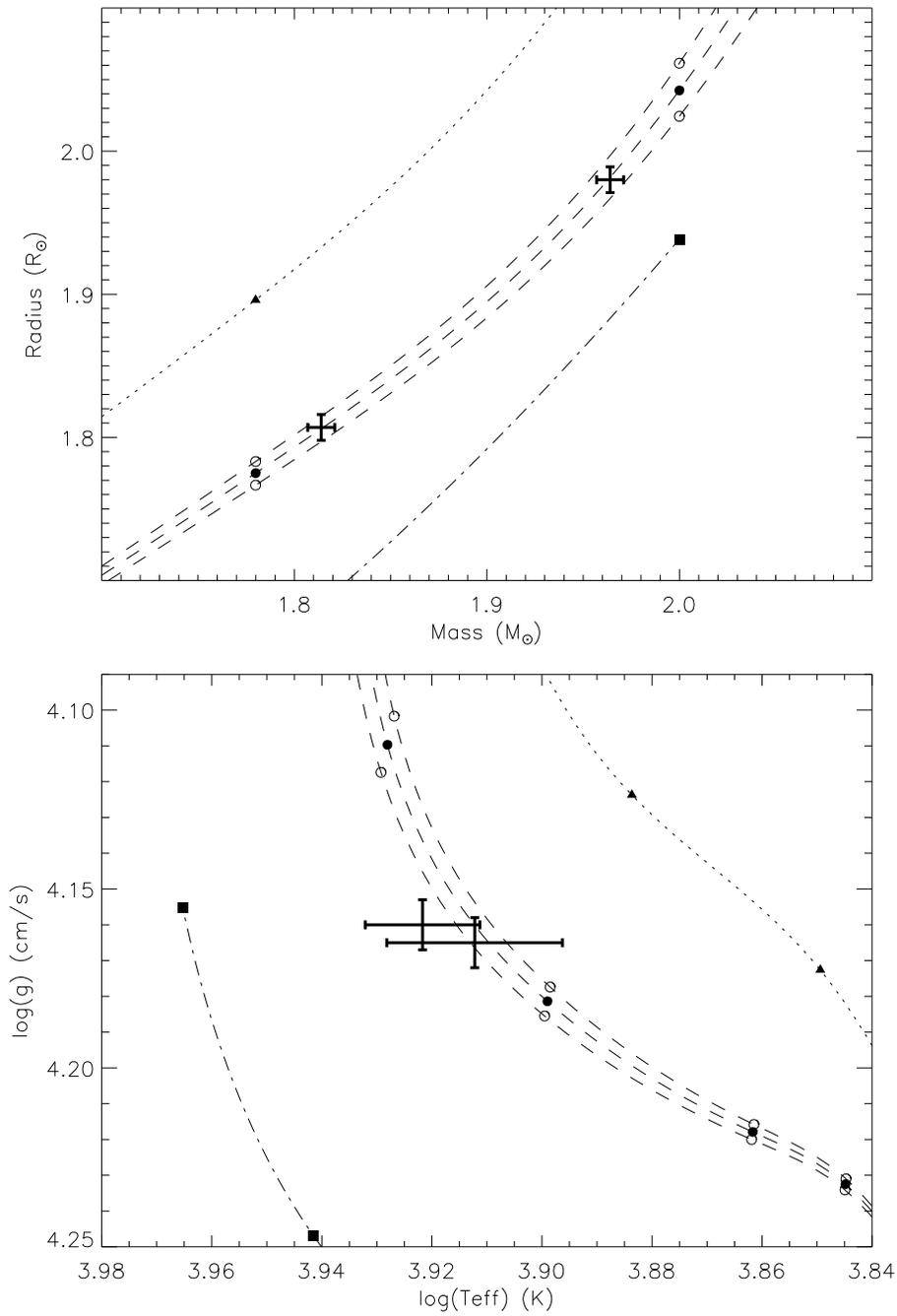} \\ 
\caption{\label{fig:wwaur:modelfit} Comparison between the astrophysical parameters of the components of WW\,Aur and the Granada theoretical stellar evolutionary models (Claret, 1995). The predictions for metal abundances of $Z = 0.01$, 0.02 and 0.03 are plotted using squares, circles and triangles, respectively. The isochrones have been plotted by interpolation in mass for all other quantities.} \end{figure}

The absolute dimensions of the two stars have been calculated from the results of the photometric and spectroscopic analyses and are given in Table~\ref{table:wwaur:dimensions}. The Granada theoretical stellar evolutionary models (Claret, 1995) are able to match the astrophysical parameters of the components of WW\,Aur, for a solar chemical composition, for an age of $565 \pm 15$\,Myr. The available predictions for other chemical compositions (Claret \& Gim\'enez, 1995; 1998; Claret, 1995; 1997) are unable to simultaneously fit the masses and radii of the two stars. Further details can be found in Southworth {\it et al}.\ (2004d).


\section{HD 23642}

\subsection{Spectroscopic analysis}

\begin{figure} \includegraphics[width=\textwidth,angle=0]{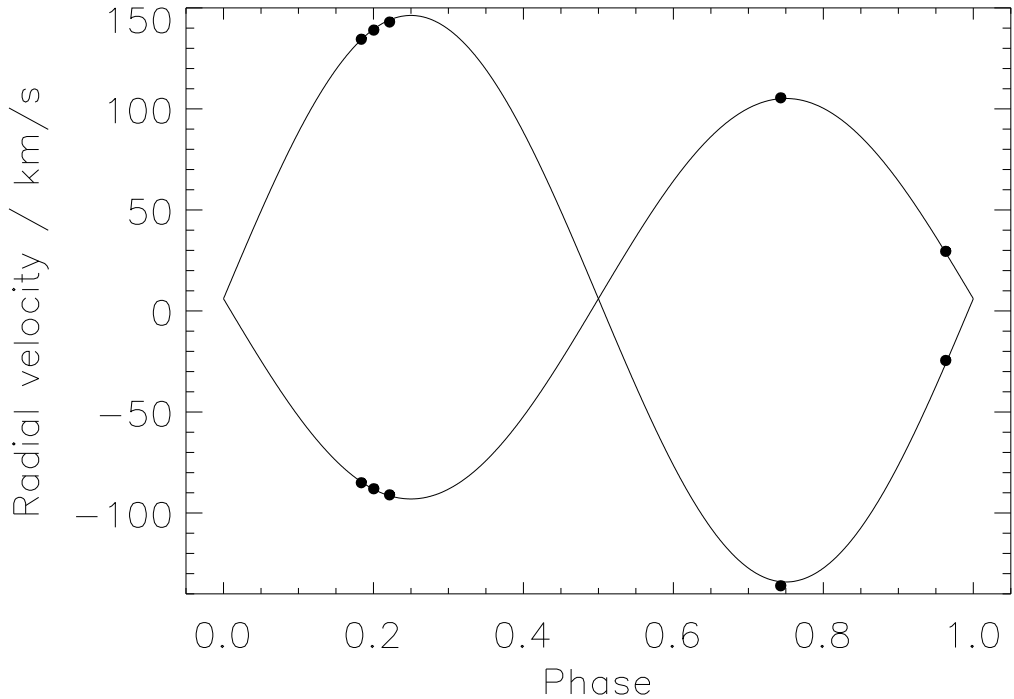} \\ \caption{\label{fig:23642:rvorbit} 
Circular spectroscopic orbit for HD\,23642 derived using the \'Elodie radial velocities. The observed radial velocities are shown with filled circles and the best-fitting spectroscopic orbit is shown using unbroken lines.} \end{figure}

A spectroscopic orbit was fitted to the five M04 \'echelle radial velocities for each star. The parameters of the orbit are $K_{\rm A} = 99.10 \pm 0.58$\kms, $K_{\rm B} = 140.20 \pm 0.57$\kms, giving $q = 0.7068 \pm 0.0052$, $M_{\rm A} \sin^3 i = 2.047 \pm 0.016$\Msun\ and $M_{\rm B} \sin^3 i = 1.447 \pm 0.013$\Msun. The orbital solution, plotted in Figure~\ref{fig:23642:rvorbit}, is in reasonable agreement with those of Torres (2003), Abt (1958) and M04.

From comparison between the observed spectra of M04 and synthetic spectra calculated using {\sc uclsyn} and {\sc atlas9} model atmospheres (see Southworth {\it et al}., 2004a for references), the effective temperatures of the two stars were found to be $9750 \pm 250$\,K and $7600 \pm 400$\,K, in good agreement with the results of M04. The uncertainties in these values include possible systematic errors caused by the slight spectral peculiarity of both components of HD\,23642.

\subsection{Light curve analysis}

\begin{figure} \includegraphics[width=\textwidth,angle=0]{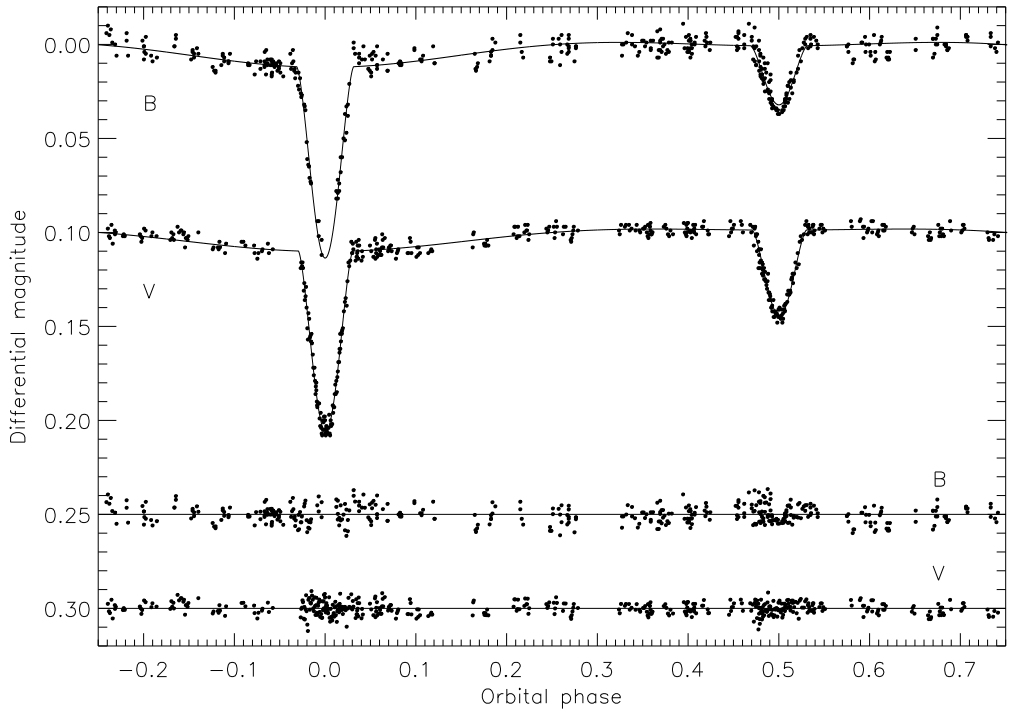} \\ \caption{\label{fig:23642:lcplot} 
The M04 $B$ and $V$ light curves with our best fitted overplotted. The $V$ light curve is shifted by +0.1\,mag for clarity. The residuals of the fit are offset by +0.25\,mag and +0.30\,mag for the $B$ and $V$ light curves respectively. Note that the poor fit around the secondary eclipse in the $B$ light curve is due to scattered data.} \end{figure}

The $B$ and $V$ light curves of M04 were solved individually using {\sc ebop}. Although these light curves are of reasonable quality, deriving accurate parameters from them is problematic due to the shallow eclipses. As contaminating `third' light, $L_3$, is poorly constrained by the observations, we have made separate solutions for $L_3 = 0.0$ and 0.05 (in units of the total light of the eclipsing stars) and included differences in the parameter values derived in the uncertainties quote below. As there are no features in the spectra of HD\,23642 known to come from a third star, it is unlikely that $L_3$ is greater than 5\%.

The ratio of the radii, $k$, is also poorly constrained by the light curves; values between 0.8 and 1.1 fit the observations equally well. To avoid this problem we have chosen to use the spectroscopic light ratio of Torres (2003), $\frac{l_{\rm B}}{l_{\rm A}} = 0.31 \pm 0.03$, found at 5187\,\AA. This has been converted to a $V$ filter light ratio using synthetic spectra, calculated from {\sc atlas9} model atmospheres, for the effective temperatures and surface gravities found in preliminary analyses. The resulting $V$ filter light ratio of $0.335 \pm 0.035$ (where the uncertainties include a minor contribution due to possible systematic errors from the use of {\sc atlas9} model atmospheres) has been used to constrain $k$ using the $V$ light curve. The resulting values of $k$ were then adopted for solution of the $B$ light curve. 

The uncertainties in the fitted parameters were again estimated using Monte Carlo simulations, including significant perturbation of the linear limb darkeing coefficients for both stars. The best fit is plotted in Figure~\ref{fig:23642:lcplot}.

\subsection{Absolute dimensions and comparison with stellar models}

The absolute dimensions of the two stars have been calculated from the results of the photometric and spectroscopic analyses and are given in Table~\ref{table:23642:dimensions}. Our results are in reasonable agreement with the results of M04, and although the radius of the secondary star is somewhat larger, the two values are consistent within their uncertainties.

\begin{table} \begin{center} 
\begin{tabular}{l r@{\,$\pm$\,}l c r@{\,$\pm$\,}l} \hline
\hspace{200pt}              & \multicolumn{2}{c}{Primary star} & \hspace{40pt} & \multicolumn{2}{c}{Secondary star} \\ \hline
Mass (\Msun)                              &    2.193 &   0.017 & &    1.550 &   0.014  \\
Radius (\Rsun)                            &    1.831 &   0.030 & &    1.548 &   0.045  \\
Surface gravity \logg\ (\cms)             &    4.254 &   0.018 & &    4.249 &   0.029  \\
Effective temperature (K)                 & 9750     & 250     & & 7600     & 400      \\
Equatorial rotational velocity (\kms)     &   38     &  1      & &   32     &   2      \\
Synchronous velocity (\kms)               &   27.7   &   0.6   & &   31.8   &   0.9    \\
\hline \end{tabular} \end{center} \caption{\label{table:23642:dimensions} Absolute dimensions and astrophysical parameters for the components of HD\,23642.} \end{table}

\begin{figure} \includegraphics[width=\textwidth,angle=0]{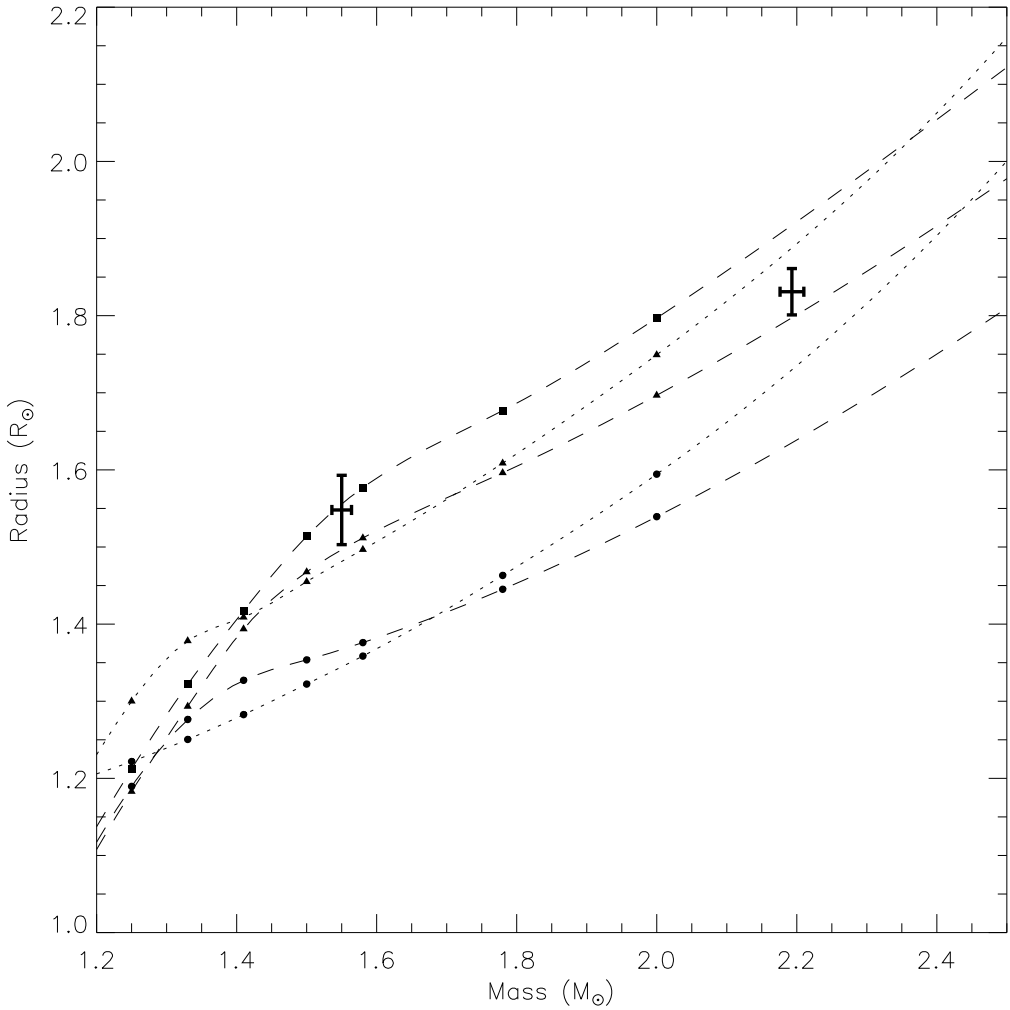} \\ \caption{\label{fig:23642:modelplot} 
Comparison between the observed properties of HD\,23642 and the Granada models for metal abundances of $Z=0.01$ (circles), $Z=0.02$ (triangles) and $Z=0.03$ (squares). Predictions for normal helium abundances are plotted with dashed lines and helium-rich model predictions ($Y = 0.34$ and 0.38 for $Z = 0.01$ and 0.02 respectively) are plotted using dotted lines. An age of 125\,Myr was assumed.} \end{figure}

In Figure~\ref{fig:23642:modelplot} the observed properties of HD\,23642 have been compared with predictions of the Granada stellar evolutionary models for $Z=0.01, 0.02$ and 0.03, adopting an age of 125\,Myr for the Pleiades (Stauffer, Schultz \& Kirkpatrick, 1998). The positions of the components of HD\,23642 in the mass--radius plane are consistent with the metal abundance of the Pleiades being solar or slightly supersolar. This confirms the atmospheric solar iron abundance determined by Boesgaard \& Friel (1990) from high-resolution spectroscopy of F dwarfs, but is in disagreement with suggestions that the `short' and `long' Pleiades distances could be reconciled by adopting a low metal abundance for the cluster.

The components of HD\,23642 cannot be fitted in the mass--radius plane by models of metal abundance or helium abundance significantly different from the solar abundances; for a solar chemical composition the predictions of the Granada stellar evolutionary models are consistent with the dimensions of HD\,23642 for ages between 125 and 175\,Myr.

\section{The Am star phenomenon}

\begin{figure} \includegraphics[width=\textwidth,angle=0]{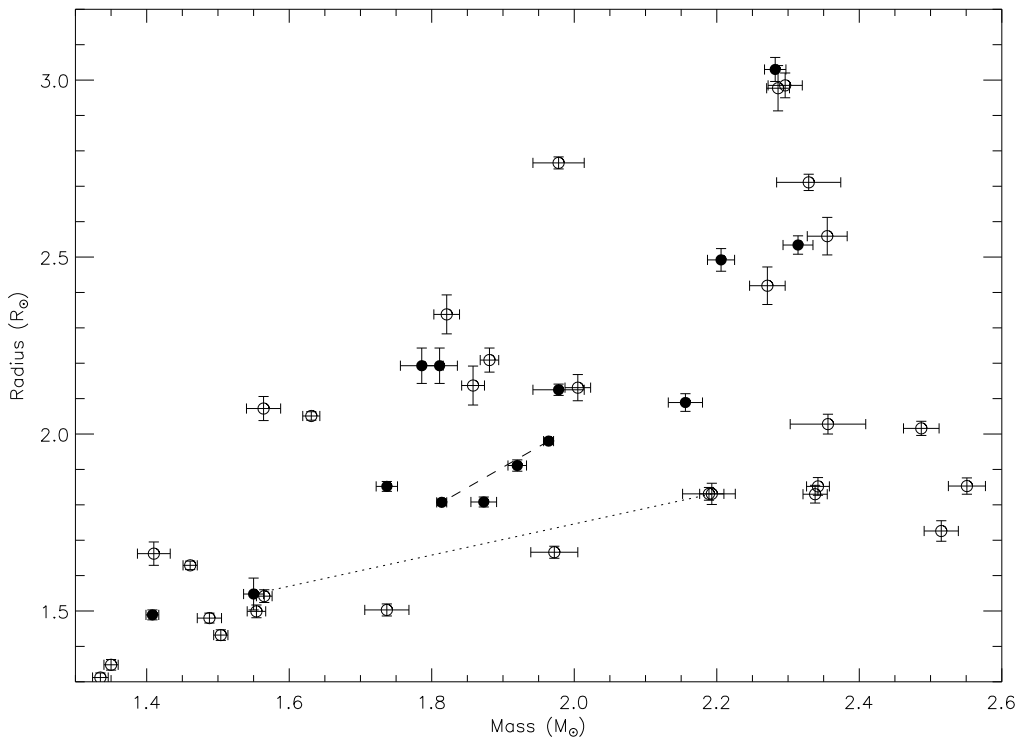}
\caption{\label{fig:MRplot} The mass--radius diagram for A-type dEB components with metallic-lined characteristics (filled circles) compared to other A-type dEB stars(open circles). The component stars of WW\,Aur are connected by a dashed line and the components of HD\,23642 are connected by a dotted line.} \end{figure}

\begin{figure} \includegraphics[width=\textwidth,angle=0]{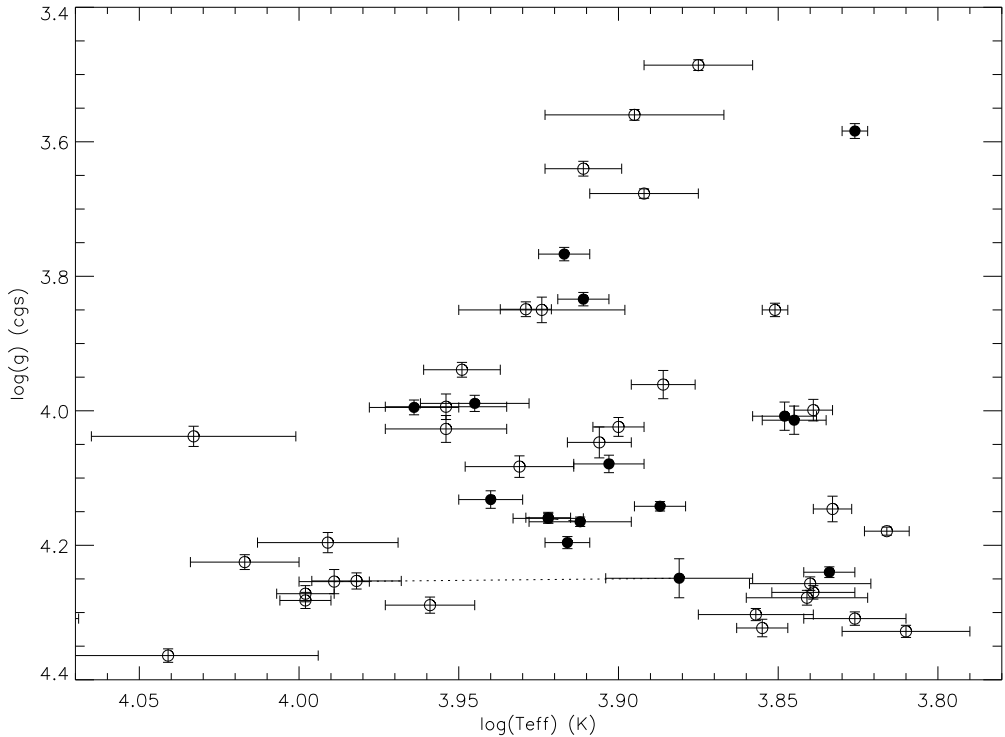}
\caption{\label{fig:TGplot} The logarithmic effective temperature--surface gravity diagram for the sample of A-type dEB stars displayed in Figure~\ref{fig:MRplot}. Symbols have the same meaning as in Figure~\ref{fig:MRplot}.} \end{figure}

\begin{figure} \includegraphics[width=\textwidth,angle=0]{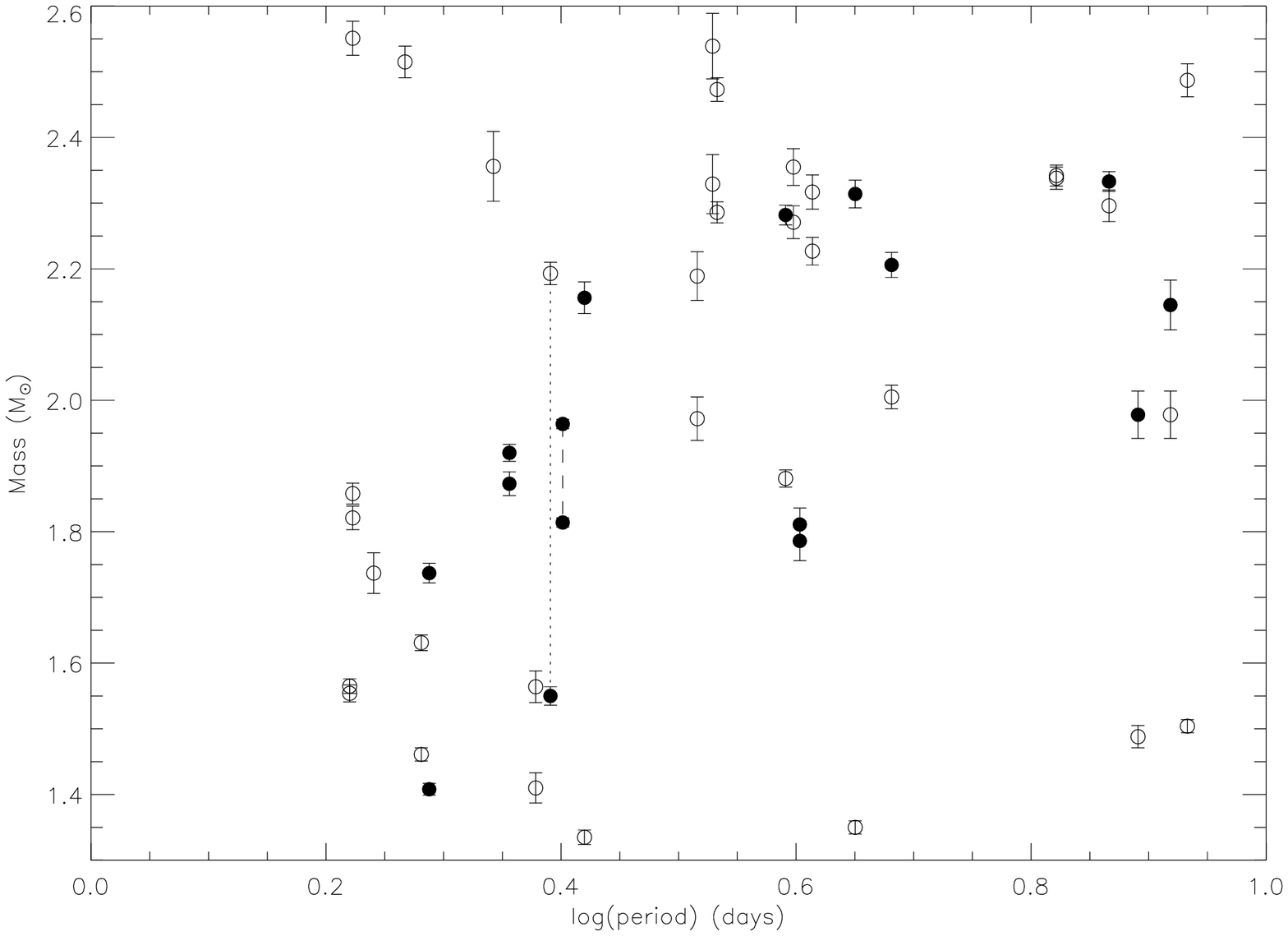}
\caption{\label{fig:PMplot} The period--mass diagram for the sample of A-type dEB stars displayed in Figure~\ref{fig:MRplot}. Symbols have the same meaning as in Figure~\ref{fig:MRplot}.} \end{figure}

\begin{figure} \includegraphics[width=\textwidth,angle=0]{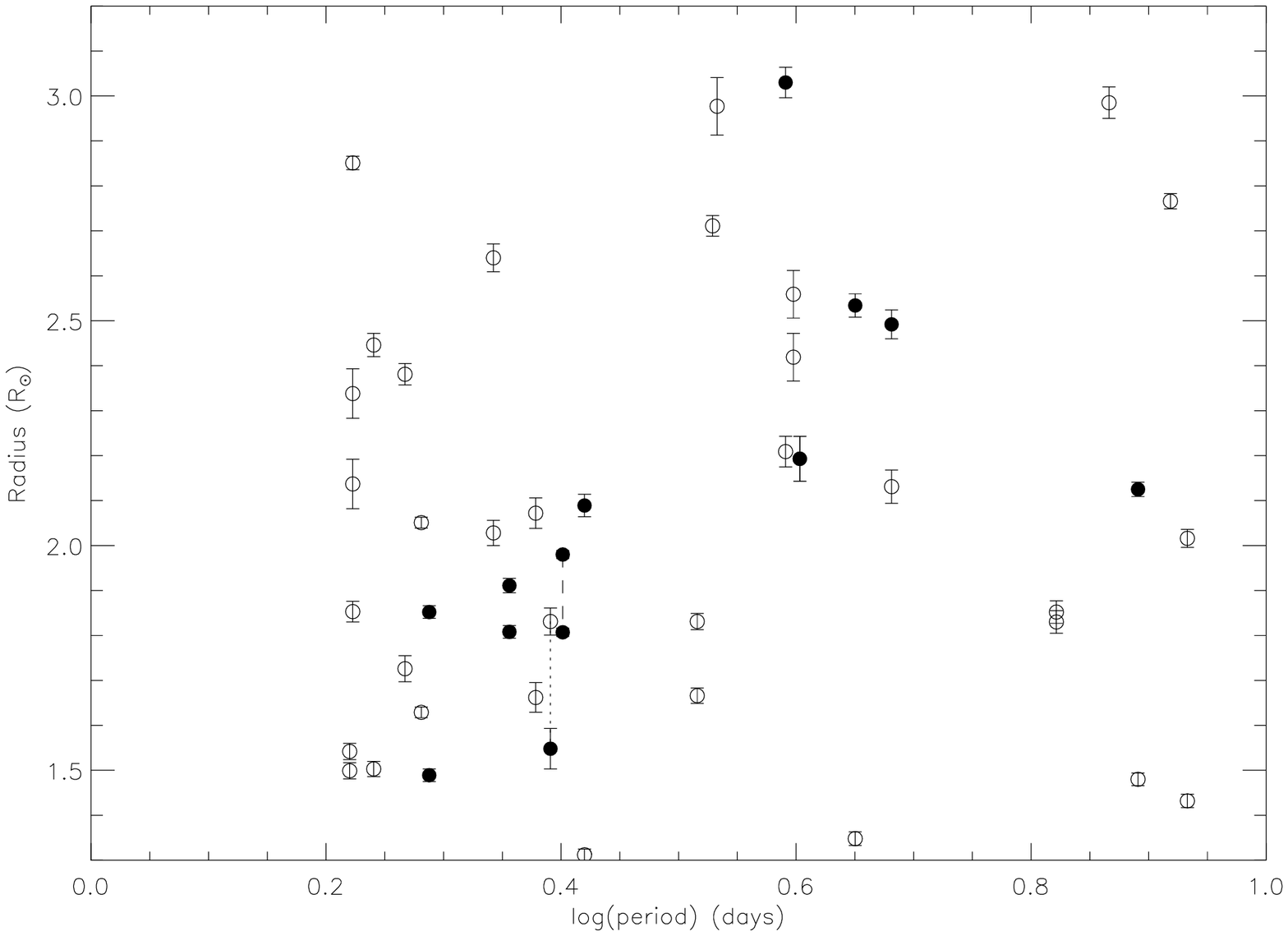}
\caption{\label{fig:PRplot} The period--mass diagram for the sample of A-type dEB stars displayed in Figure~\ref{fig:MRplot}. Symbols have the same meaning as in Figure~\ref{fig:MRplot}.} \end{figure}

There are a significant number of dEBs with components between 1.5 and 2.5\Msun\ which have accurate masses and radii. We have taken a list from Andersen (1991) and added eight systems which have been studied more recently (EI\,Cep, WW\,Cam, BP\,Vul, GG\,Ori, FS\,Mon, V364\,Lac and WW\,Aur and HD\,23642). These stars have been plotted in mass--radius, temperature--gravity, period--mass and period--radius plots (Figures \ref{fig:MRplot}, \ref{fig:TGplot}, \ref{fig:PMplot} and \ref{fig:PRplot}).

It can be seen from the figures above that there is no region of astrophysical parameter space for which all stars are  metallic-lined, although there is a cluster of Am stars around mass $M = 1.9$\Msun\ and surface gravity $\logg = 4.15$ (\cms) which includes both components of WW\,Aur. Suggestions that metallic-lined stars have slightly larger radii than expected for their spectral type (Budaj, 1996) are therefore not confirmed. A-type stars of higher mass, though,  do seem to need a lower surface gravity to be Am stars.

As the predictions of theoretical models of approximately solar chemical composition can successfully match the masses and radii of both HD\,23642 and WW\,Aur, the metallic-lined spectrum phenomenon is confirmed to be a surface feature which does not significantly modify the interior conditions of the star. It is also clearly not a reflection of an overall high metal abundance. It remains unclear why some stars display metallic spectral lines when many similar ones do not.

It should be remembered that using well-studied dEBs to investigate phenomena, such as enhanced metallic lines, can be misleading. The observational sample is very biased towards slightly evolved stars (Andersen, 1991) because these dEBs are more likely to show eclipses. A bias also exists towards dEBs with mass ratios close to unity because the observational requirements of such systems are more easily satisfied.


\section{The distance to the Pleiades}

\subsection{Distance using bolometric corrections}

Adopting the astrophysical parameters of HD\,23642 from Table~\ref{table:23642:dimensions} and using bolometric corrections from Bessell, Castelli \& Plez (1998) to convert absolte bolometric magnitudes into absolute visual magnitudes, we find a distance to HD\,23642 of $138.1 \pm 4.7$\,pc. Whilst this is an acceptable result to adopt for the final distance to HD\,23642, it is still affected by (probably small) systematic error due to the use of bolometric corrections calculated from theoretical models.

The distance found by M04, $131.9 \pm 1.9$\,pc, was derived using the above method but is significantly different from the result we obtain. The probable reason for this, which is consistent with the results given by M04, is that the absolute bolometric magnitudes given by M04 were calculated using values of the solar absolute bolometric magnitude and luminosity of $\Mbolsun = 4.77$ and $\Lsun = \sci{3.906}{26}$\,W, whereas the bolometric corrections used (Bessell {\it et al.}, 1998) were calculated with $\Mbolsun = 4.74$ and $\Lsun = \sci{3.855}{26}$\,W. This inconsistency is sufficient to explain the difference between the distance derived by M04 and by ourselves, from the same observational data.

\subsection{Distance using surface brightness calibrations}

An alternative to considering the unobserved parts of the stellar spectral energy distributions is to use empirical relations between effective temperature and the wavelength-dependent surface brightness of a star. In this method the angular diameter of the star is compared to its linear diameter, determined from photometric analysis, to find the distance. The main source of uncertainty in this method is usually the light ratio, found from the photometric analysis, needed to calculate the individual magnitudes of the components of the dEB from the apparent magnitude of the system. If the effective temperatures of the two stars are known independently, this difficulty can be bypassed by comparing the apparent magnitude of the system to the flux-weighted combined angular diameter of the two stars.

From consideration of the definitions of surface brightness and angular diameter, it can be shown that the distance to a dEB is given by
\begin{equation} \label{eq:d}
d = 10^{0.2 m_\lambda} \sqrt{\left[\frac{2 R_{\rm A}}{\phi_{\rm A}^{(m_\lambda = 0)}}\right]^2 + 
                             \left[\frac{2 R_{\rm B}}{\phi_{\rm B}^{(m_\lambda = 0)}}\right]^2 }
\end{equation}
where, for a distance given in parsecs, the stellar absolute radii,  $R_{\rm A}$ and $R_{\rm B}$, are given in AU, the zeroth-magnitude angular diameters $\phi_{\rm A}^{(m_\lambda = 0)}$ and $\phi_{\rm B}^{(m_\lambda = 0)}$ are given in arcseconds and $\lambda$ represents a broad-band filter passband. Calibrations for $\phi^{(m_\lambda = 0)}$ are given in terms of effective temperature by Kervella {\it et al}.\ (2004). The scatter of the fits is smaller at infrared wavelengths. We have applied the calibrations for the $B$ and $V$ filters (using Tycho apparent magnitudes transformed to the Johnson system) and for the $JHK$ filters (using 2MASS data transformed to the SAAO system). We adopt the result from the $K$ filter calibration, a distance of $139.1 \pm 3.6$\,pc, as our final distance to HD\,23642 and so to the Pleiades. The main source of uncertainty in this distance is from the uncertainty in the apparent $K$ magnitude of HD\,23642.


\section{Conclusion}

Preliminary results have been given from analysis of the A-type dEB WW\,Aurigae, which is composed of two metallic-lined stars. The masses have been derived to accuracies of 0.3\% and the radii to accuracies of 0.5\%; predictions of the Granada theoretical stellar evolutionary models agree with the dimensions of the stars for an approximately solar chemical composition and an age of $565 \pm 15$\,Myr. However, the agreement between the $uvby$ and $UBV$ light curves and the light curve of Huffer \& Kopal is not good. If the latter data are rejected, the ratio of the radii increases to $0.953 \pm 0.011$, and the masses and radii of the two stars are unable to be fitted by theoretical models for one single age. For this reason, the study of WW\,Aurigae presented here must be regarded as preliminary.

HD\,23642 is an A-type dEB which is a member of the Pleiades open cluster, for which the distance is currently controversial. The data of M04 have been reanalysed to provide robust parameter values and uncertainties. The metal abundance of the Pleiades has been estimated by comparing the masses and radii of the component stars of HD\,23642 with the predictions of theoretical stellar models. The resulting metal abundance, approximately solar, disagrees with suggestions that the `short' and `long' distances to the Pleiades can be reconciled by adopting a low metal abundance for the cluster. The distance to HD\,23642 has been estimated to be $139 \pm 4$\,pc from the use of surface brightness--effective temperature relations provided by Kervella {\it et al}.\ (2004). This distance is in agreement with the `long' distance to the Pleiades and in disagreement with the Hipparcos distance to HD\,23642 and the Pleiades.

\begin{acknowledgments}
The authors would like to thank Dr.\ U.\ Munari for extremely frank discussions and for making his data on HD\,23642 freely available.

JS acknowledges financial support from PPARC in the form of a postgraduate studentship. The authors acknowledge the data analysis facilities provided by the Starlink Project which is run by CCLRC on behalf of PPARC. The following internet-based resources were used in research for this paper: the NASA Astrophysics Data System; the SIMBAD database operated at CDS, Strasbourg, France; the VizieR service operated at CDS, Strasbourg, France; and the ar$\chi$iv scientific paper preprint service operated by Cornell University. 

This publication makes use of data products from the Two Micron All Sky Survey, which is a joint project of the University of Massachusetts and the Infrared Processing and Analysis Center/California Institute of Technology, funded by the National Aeronautics and Space Administration and the National Science Foundation. 
\end{acknowledgments}

\end{document}